\documentclass{article}
\usepackage{ijcai25}

\usepackage{times}
\usepackage{soul}
\usepackage{url}
\usepackage[hidelinks]{hyperref}
\usepackage[utf8]{inputenc}
\usepackage[small]{caption}
\usepackage{graphicx}
\usepackage{amsmath}
\usepackage{amsthm}
\usepackage{booktabs}
\usepackage{algorithm}
\usepackage{algorithmic}
\usepackage[switch]{lineno}
\usepackage{layout}  % 导入layout包
\usepackage{enumitem} % 需要先导入这个宏包
\usepackage{xcolor}
\usepackage{makecell}
\usepackage{bm}
\usepackage{multirow}

\title{Celler: A Genomic Language Model for Long-Tailed Single-Cell Annotation}

\author{
Huan Zhao$^{1*}$
\and
Yiming Liu$^{1,2}$\and
Jina Yao$^{1,2}$\and
Ling Xiong$^{1,2}$\and
Zexin Zhou$^{1,2}$\and
Zixing Zhang$^{1*}$\\
\affiliations
 $^1$School of Computer Science and Electronic Engineering, Hunan University, Changsha 410082, China\\
 $^2$Higentec AiLab\\
\emails
\{hzhao@hnu.edu.cn,
zixingzhang@hnu.edu.cn
}

\begin{document}
% \layout
\maketitle

\begin{abstract}

% 单细胞技术的最新进展为揭示复杂生物系统的分子复杂性带来了前所未有的机会，尤其是那些与人类特异性疾病相关的生物系统。这一进展同时带来了一个新的挑战——与疾病状况相关的长尾单细胞数据的注释。为了有效地应对这一挑战，我们构思了Celler，这是一种前卫的生成预训练模型，专门为单细胞数据注释设计。Celler整合了两个开创性的元素：首先，我们引入了高斯膨胀（GInf）损失。通过动态调整样本权重，GInf Loss显著增强了模型从罕见类别中学习的能力，同时减少了普遍类别的过拟合风险。其次，我们将硬数据挖掘（HDM）策略注入到我们的训练方法中，这大大提高了模型的预测准确性。全面的实验验证强调Celler在管理复杂数据集方面的卓越能力，为单细胞研究领域提供前所未有的精度和可扩展性融合。

% 单细胞技术的最新进展为揭示复杂生物系统的分子复杂性带来了前所未有的机会，尤其是那些与人类特异性疾病相关的生物系统。然而，这一进展也带来了新的挑战——如何对与疾病相关的长尾单细胞数据进行高效注释。为了有效应对这一挑战，我们提出了Celler，这是一种前沿的生成预训练模型，专门为单细胞数据注释而设计。Celler整合了两个开创性的元素：首先，我们引入了高斯膨胀（GInf）损失函数。通过动态调整样本权重，GInf Loss显著增强了模型在罕见类别上的学习能力，同时减少了常见类别的过拟合风险。其次，我们在训练过程中融入了困难样本挖掘（HDM）策略，从而大幅提升了模型的预测准确性。

% 此外，为了更深入地探索这一领域，我们构建了一个涵盖80个人体组织和75种特异性疾病、包含4000万个细胞的大规模单细胞数据集——Celler-75。该数据集为全面研究单细胞技术在疾病研究中的潜力提供了重要支撑。
Recent breakthroughs in single-cell technology have ushered in unparalleled opportunities to decode the molecular intricacy of intricate biological systems, especially those linked to diseases unique to humans. However, these progressions have also ushered in novel obstacles—specifically, the efficient annotation of extensive, long-tailed single-cell data pertaining to disease conditions. To effectively surmount this challenge, we introduce Celler, a state-of-the-art generative pre-training model crafted specifically for the annotation of single-cell data. Celler incorporates two groundbreaking elements: First, we introduced the Gaussian Inflation (GInf) Loss function. By dynamically adjusting sample weights, GInf Loss significantly enhances the model’s ability to learn from rare categories while reducing the risk of overfitting for common categories. Secondly, we introduce an innovative Hard Data Mining (HDM) strategy into the training process, specifically targeting the challenging-to-learn minority data samples, which significantly improved the model's predictive accuracy. Additionally, to further advance research in this field, we have constructed a large-scale single-cell dataset: Celler-75, which encompasses 40 million cells distributed across 80 human tissues and 75 specific diseases. This dataset provides critical support for comprehensively exploring the potential of single-cell technology in disease research. Our code is available at https://github.com/AI4science-ym/HiCeller.

\end{abstract}

%================================================================

%第一章
%介绍
%（第一段）任务介绍+首先介绍单细胞技术的先进性优越性（褒），并指出人工繁重的问题（贬）。

\begin{figure}[t] % [t]表示将浮动图片放置在页面顶部
    \vspace*{-5mm} % 调整图片与上方的距离
    %\hspace*{\fill} % 将图片推到右侧
    \includegraphics[width=0.5\textwidth]{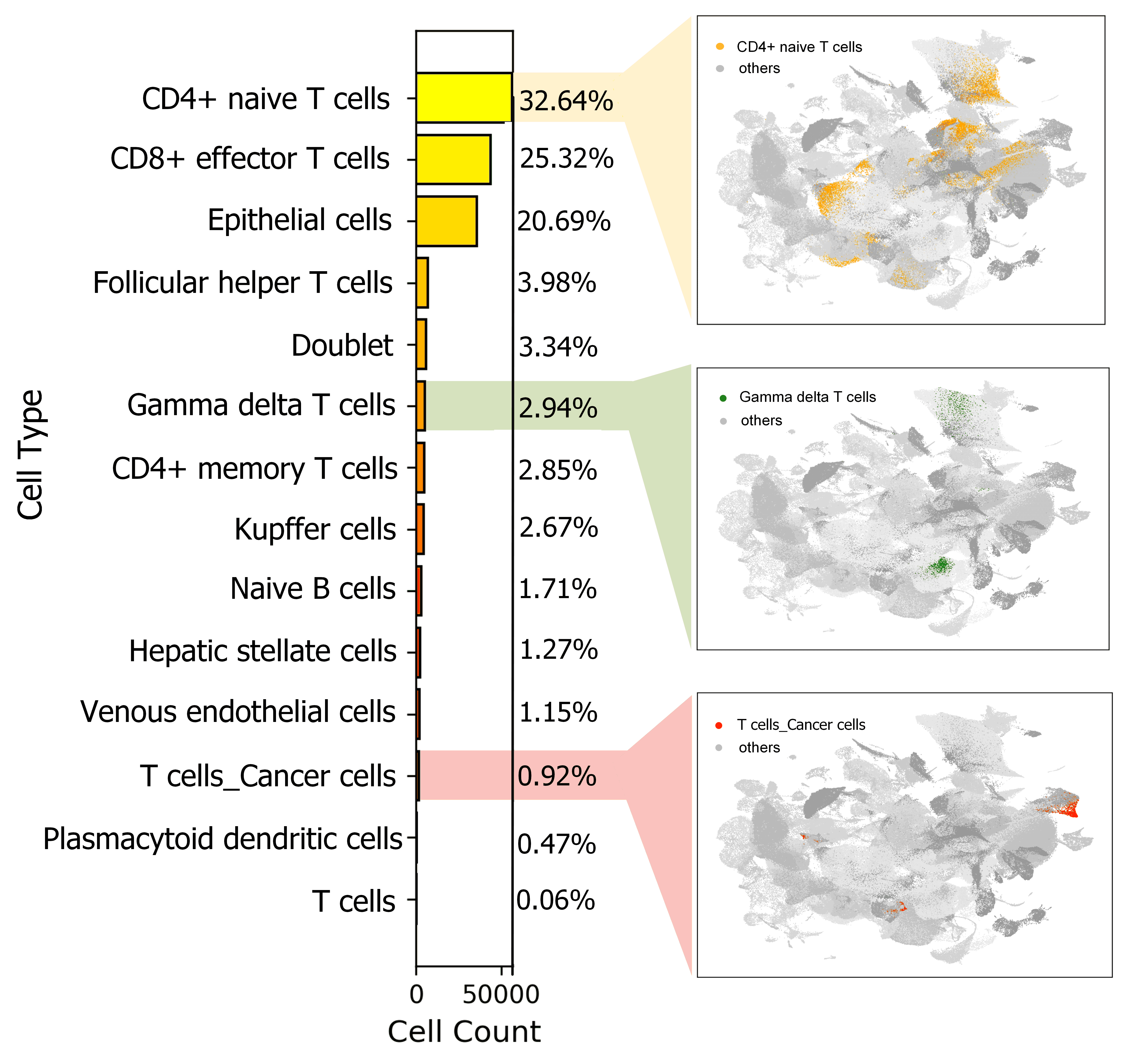} % 插入图片，并设置宽度为50%页面宽度
    \caption{Long-Tail Data Distribution} % 设置图片标题
    \label{fig:long_tail} % 设置图片标签，便于引用
    \vspace*{-5mm} % 调整图片与上方的距离
\end{figure}
\section{Introduction}

% 单细胞RNA测序（scRNA-seq）是一种通过研究单细胞水平的细胞内转录物（mRNA）表达，揭示不同细胞间基因表达差异和细胞功能异质性的技术。单细胞RNA测序（scRNA-seq）技术的快速发展显著提升了我们对细胞异质性和疾病机制的理解，使得在单细胞水平上精确表征不同细胞类型成为可能.\cite{plass2018cell}\cite{cao2019single}\cite{schaum2018single}\cite{zhao2020evaluation}。与传统的群体水平转录组测序相比，单细胞转录组测序能够获得每个细胞更为精确的基因表达信息，从而解决了群体水平测序中因细胞混杂导致的平均化效应问题。然而，单细胞数据规模的爆炸式增长也带来了显著挑战，尤其是在数据注释方面。在处理大规模的数据时，传统的手动注释方法由于依赖人工操作，不仅耗时费力，手动方法尤其繁琐且容易出错。
Single-cell RNA sequencing (scRNA-seq) is a technique that reveals differences in gene expression and cellular functional heterogeneity among different cells by studying intracellular transcript (mRNA) expression at the single-cell level \cite{grabski2022probabilistic}. The rapid development of single-cell RNA sequencing (scRNA-seq) technology has greatly improved our understanding of cellular heterogeneity and disease mechanisms \cite{lukassen2020sarscov2,he2020single,mcdavid2013data}, making it possible to accurately characterise different cell types at the single-cell level \cite{plass2018cell,cao2019single,schaum2018single,zhao2020evaluation,pliner2019supervised}. Compared with traditional cluster-level transcriptome sequencing, single-cell transcriptome sequencing provides more accurate gene expression information for each cell, thus solving the problem of averaging effect due to cell mixing in cluster-level sequencing \cite{huang2020evaluation,tarashansky2021mapping}. However, the explosion in the amount of single-cell data also poses significant challenges, especially in data annotation. When dealing with large-scale data, traditional manual annotation methods are not only time-consuming and labour-intensive due to their reliance on manual operations, but also the manual methods are particularly cumbersome and error-prone.

% 面对mRNA数据的高维特性，传统的机器学习方法在降维过程中只能发现数据中的线性特征。基因对于细胞表达的意义就像单词对句子表达的意义一样，大型语言模型通过学习上下文关系来捕捉单词的嵌入表示，同样，由基因构成的复杂网络也可以通过大型语言模型来学习基因之间的相互关系。例如，一些更具前瞻性的基因组语言模型(GLM)，如scBERT（Yang等，2022），scGPT（Cui等，2023） scPLM()，Geneformer（Theodoris等，2023）和tGPT（Shen等，2023），这些都是专门为单细胞数据设计的预训练模型。这些模型将基因视为词元(tokens)，随机屏蔽部分非零基因表达值，并根据剩余数据进行预测，从而有效地捕捉了基因之间的复杂关系，进而提升了细胞表示的能力。

Due to the high dimensionality of mRNA data, traditional machine learning methods are typically limited to capturing linear features during the dimensionality reduction process, making it challenging to account for the intrinsic associations within RNA expression data and the nonlinear expression characteristics across different cells \cite{pasquini2021automated}. To address this limitation, researchers in recent years have introduced large language models, aligning the conceptual framework of transcriptomic data with that of natural language data \cite{ganin2015unsupervised,zhang2010understanding}. For example, some state-of-the-art genomic language models (GLMs), such as scBERT \cite{scbert}, scGPT \cite{scgpt}, CellPLM \cite{ICLR2024cellplm}, Geneformer \cite{theodoris2023geneformer} and tGPT \cite{shen2023tgpt}, are specifically designed for single-cell data designed as pre-trained models. These models treat genes as tokens (words), randomly mask some non-zero gene expression values and make predictions based on the remaining data, thus effectively capturing complex relationships between genes and improving cell representation.

% 然而上述方法在深入理解这些疾病的分子机制过程中出现了一定的疲态。这是因为病变相关细胞的数量通常显著少于正常组织细胞，导致数据呈现出明显的长尾分布特征，尤其是在癌症等复杂疾病的研究中，某些关键细胞或基因可能表现出代表性不足，这使得模型难以有效学习重要信息，如图1所示在患有肺癌的肺部器官的细胞表观，其肺癌细胞占比仅有0.92%。在这种情况下，上述模型训练中对病变细胞的关注可能会被大量正常细胞所稀释，进而削弱对这些关键细胞的识别能力，甚至可能扭曲我们对疾病机制的理解。
Nevertheless, the methods outlined earlier exhibit certain limitations in deeply deciphering the molecular mechanisms of these diseases. This is mainly due to the fact that the number of disease-related cells is typically vastly outnumbered by normal tissue cells, causing the data to exhibit marked long-tail distribution characteristics. This is particularly pronounced in the investigation of complex diseases such as cancer, where certain pivotal cells or genes may be underrepresented, impeding the model's ability to effectively apprehend critical information. As shown in Figure \ref{fig:long_tail}, in the cell appearance of the lung organ of a patient with lung cancer, the proportion of lung cancer cells is only 0.92\%. In this case, the attention to diseased cells in the above model training may be diluted by a large number of normal cells, thereby weakening the recognition ability of these key cells, and even possibly distorting our understanding of disease mechanisms.

% 为了解决这一问题，我们引入了 Gaussian Inflation (GInf) Loss，这是一种专为长尾数据设计的损失函数，旨在增强模型对稀有类别的敏感性。GInf Loss通过在动态空间中，对依据类别数据量的大小，动态的为尾部类别以高斯分布的模式对每个数据实例的特征进行权值的扩增，从而良性的调整正负样本在特征空间所占的比重，有效缓解了模型在处理稀有细胞类型时的学习缺陷，从而使网络能够更加聚焦于病变细胞及其相关基因。不仅如此，我们额外提出了一种Hard Data Mining(HDM)困难样本挖掘训练策略，这种策略通过将模型最终输出的分类置信度作为评判指标，通过HDM把错误的预测了类别且置信度很高的数据定义为困难样本，在训练过程中给予困难样本数据额外的关注，增加困难数据的训练轮次，从而提升模型的准确度。在深度探索这一领域的过程中，我们不仅筹集，而且构建了一个大规模的私有数据集Celler-75，其无与伦比的数据量达到了4千万个已标注的细胞数据，覆盖了80种人体组织和75种特异性疾病。这个数据集的规模、深度和广度，都大大超过了目前任何公开可用的数据集。以公开数据集Multiple Sclerosis（MS）和human pancreas（hPancreas）为例，它们的数据量相形见绌。MS数据集的训练集只含1.3万个细胞样本，测试集仅有7千个；同时，hPancreas数据集的训练集仅有1万个细胞样本，测试集更是只有4千个。
To address this issue, we propose Gaussian Inflation (GInf) Loss, a loss function specifically designed for long-tailed data, which aims to enhance the model's sensitivity to rare categories. GInf Loss dynamically increases the feature weights of individual data instances from tail categories in a Gaussian distribution pattern, based on the size of the category data in the dynamic space. This approach benignly adjusts the balance between positive and negative sample weights in the feature space, effectively mitigating the model's learning deficiencies in handling rare cell types. As a result, the network is better able to focus on diseased cells and their associated genes.

In addition, we propose a Hard Data Mining (HDM) training strategy for difficult sample mining. This strategy utilizes the model's final output classification confidence as an evaluation metric, defining misclassified samples with high confidence as difficult samples. During training, additional attention is given to these difficult samples by increasing their training iterations, thereby enhancing the overall accuracy of the model. 

In the course of our in-depth exploration in this field, we have not only proposed novel methods but also constructed a large-scale private dataset, Celler-75. This dataset boasts an unparalleled volume, comprising 40 million annotated cell samples, covering 80 types of human tissues and 75 specific diseases. To the best of our knowledge, the scale, depth, and breadth of this dataset far surpass any publicly available datasets at present. For instance, when compared to public datasets such as Multiple Sclerosis (MS) \cite{MSdataset} and human pancreas (hPancreas) \cite{hPancreas-dataset}, the difference in data volume becomes evident. The MS dataset contains only 13,468 cell samples in its training set and 7,000 in its test set. Similarly, the hPancreas dataset includes only 10,600 cell samples in its training set and just 4,218 in its test set.

% 我们的研究结果显示，Celler在面对公开数据集MS、hPancreas以及我们的大规模私有数据集Celler-75时，表现都极为出色。它在识别稀有细胞类型的准确度上，相较于scBert模型提高了超过10%，其整体F1 Score也显著优于其他基准模型，凸显了其卓越的泛化能力和实际应用价值。高效的单细胞数据注释对于疾病诊断、个性化医学以及生物标志物的发现具有重要意义。Celler的提出不仅打破了现有注释方法在处理长尾分布数据时的局限性，同时也为未来复杂多组学数据的整合分析提供了有力的工具支持。
% Our research results show that Celler performs extremely well when facing the public datasets MS, hPancreas, and our large-scale private dataset Celler-75. It improves the accuracy of identifying rare cell types by over 10\% compared to the scBert model, and its overall F1 Score is also significantly better than other benchmark models, highlighting its excellent generalization capability and practical application value. Efficient single-cell data annotation is of great significance for disease diagnosis, personalized medicine, and the discovery of biomarkers. The introduction of Celler not only breaks the limitations of existing annotation methods in handling long-tail distribution data but also provides strong tool support for the integrated analysis of complex multi-omics data in the future.
%总的来说，我们的贡献如下：
%（1）我们开发了一个以人类疾病组织为基础的包含75类人类疾病的单细胞数据集，其中包括4000多万个细胞和900多万个基因。
%（2)我们引入GInf Loss来解决疾病数据中的长尾分布，使模型能够专注于罕见疾病细胞类型和特定基因。
%（3）通过引入Hard Data Mining(HDM)训练策略，提高了模型整体的性能。

Overall, our contributions are listed below:
\begin{itemize}[leftmargin=10pt] % 设置缩进为10pt，可以根据需要调整
    \setlength{\itemsep}{0pt} % 调整行间距
    \item We have developed the largest known single-cell dataset based on human disease tissues, which includes 75 types of human diseases, containing over 40 million cells and over 9 million genes.
    \item We introduced GInf Loss to address the long-tail distribution in disease data, enabling the model to focus on rare disease cell types and specific genes.
    \item By introducing the Hard Data Mining (HDM) training strategy, the overall performance of the model has been improved.
\end{itemize}

\section{Related works}
\subsection{Genomic Language Model}
This work focus on applying large-scale model technology in the single-cell domain, expanding sc-RNAseq data, and designing experiments to achieve cell annotation.

A notable contribution in this field is scBERT \cite{scbert}, which employs a multi-layer Performer to pretrain scRNAseq data, followed by fine-tuning to adapt to various downstream tasks. Building on this foundation, the work by xTrimoGene \cite{gong2023xtrimogene} enhanced scBERT with two key improvements: pruning zero-expression genes and refining the expression binning strategy through automatic discretization. These modifications significantly boosted the model's scalability and feature resolution. The latest preprint, scGPT \cite{scgpt}, introduced a variant of masked language modeling that mimics autoregressive generation in natural language processing, iteratively predicting masked genes based on the model's confidence. In contrast, CellPLM \cite{ICLR2024cellplm} proposes a pretraining method for cell language models that goes beyond single cells. CellPLM not only captures the gene expression patterns within individual cells but also considers the interactions between cells and the tissue structure, thereby providing a more comprehensive understanding of cellular functions. These innovative approaches open new avenues for single-cell omics research, advancing our understanding of cellular functions.

\subsection{Long-Tail for Classification}
\textbf{Resampling and Reweighting} \quad In the real world, data often follow a long-tailed distribution, which can pose challenges to the classifier. To mitigate data imbalance, directly under or up sampling\cite{resample1,resample2,resample3,resample4} the training instance based on the relation of class size is a straightforward method which drives us to repeat the learning tail instance and ignore the learning of head classes. Reweighting assigns loss functions to the sample of different class and adjusts the effect of label frequency on the loss function. 

\textbf{Integrated Learning}\quad Ensemble learning methods have demonstrated outstanding performance in addressing long-tailed classification tasks. Researchers leverage individual expert modules to capture the distribution characteristics of different data groups, subsequently combining the learning outcomes from each expert module. During the training phase, these expert modules operate independently, avoiding mutual interference and enabling focused and efficient learning. Expert-based ensemble strategies, such as BBN \cite{BBN} and RIDE \cite{RIDE}, not only allow the model to focus more effectively on learning from tail data but also enhance its overall performance.

% 集成学习方法在解决长尾分类任务中表现出色。研究人员利用单独的专家模块来捕获不同数据组的分布特性，随后将每个专家模块的学习结果进行组合。在训练阶段，这些专家模块独立运行，避免了相互干扰，从而实现专注且高效的学习。如 BBN\cite{BBN} 和 RIDE\cite{RIDE}这种基于专家的集成策略不仅使模型能够更有效地专注于尾部数据的学习.
\section{Methods}
% 细胞注释是单细胞转录组分析的重要步骤，其目的是根据单细胞的基因表达谱来推断每个细胞的类型、状态或功能。细胞注释的核心依据主要是根据已知的细胞类型特异性表达的标志性基因（marker genes）来推断细胞类型。例如：CD3D/CD3E 用于标记 T 细胞。CD19 用于标记 B 细胞。CD68 用于标记巨噬细胞。ACTA2 用于标记成肌纤维细胞。通过差异表达分析（Differentially Expressed Genes, DEG），找到每个细胞亚群中高度特异性表达的基因，结合已知标志基因进行注释。单细胞转录组分析中，常用降维（如 PCA、t-SNE、UMAP）和聚类（如 Louvain、Leiden 算法）方法，将表达模式相似的细胞聚为一类。对每个聚类的差异表达基因进行分析，结合标志基因和数据库进行注释。

% 为了应对mRNA数据的高维度性以及错综复杂的关系，我们提出Genomic Language Model（GLM）预训练模型，在我们构建的数据集上进行无监督训练，学习到丰富的基因相互关系，同时考虑到数据的长尾分布以及不均衡性，我们加入“GINF”损失函数和HDM,具体实现细节见后续章节。

Cellular annotation is an important step in single-cell transcriptome analysis, which aims to infer the type, state, or function of each cell based on the gene expression profile of a single cell \cite{mofitt2018molecular,brbic2020mars}. Fundamentally, cellular annotation involves inferring cell types through the expression of specific marker genes \cite{cao2020scsa}. For example, CD3D/CD3E is used to label T cells, CD19 is used to label B cells, CD68 is used to label macrophages, and ACTA2 is used to label myofibroblasts. Through Differential Expressed Genes (DEG) analysis, genes that are highly specifically expressed in each cell subpopulation are found and annotated with known marker genes. For single-cell transcriptome analysis, dimensionality reduction (e.g., PCA, t-SNE, UMAP) and clustering (e.g., Louvain, Leiden's algorithm) methods are commonly used to cluster cells with similar expression patterns into one group \cite{zhang2019scina}. Differentially expressed genes in each cluster are analyzed and annotated in combination with marker genes and databases.

In order to cope with the high dimensionality and intricate relationships of mRNA data, we propose the Genomic Language Model (GLM) pre-training model, which is unsupervisly traind on our constructed dataset to learn the rich gene interrelationships, and considering the long-tailed distribution of the data and the imbalance of the data, we incorporate the ``GInf'' loss function and HDM, and the details of the implementation are shown in the following sections.

\subsection{Pretraining Process}

% 我们构建了 Genomic Language Model（GLM），并在预训练阶段将大语言模型中的数据处理概念转移到 Genomic Language Model 的框架中。具体而言，我们将单细胞转录组数据的基本单位重新定义为语言模型中的概念结构：每个基因的表达量被视为最小的数据单元，这类似于语言模型中的 token，即一个具有独立语义的基本片段。在单细胞转录组数据中，每个基因的表达量反映了其在细胞功能和状态中的独特作用，这与 token 在语言模型中承载语义的功能类似。同时，我们将由所有基因表达量构成的一个细胞整体，视为类似于语言模型中的 句子。在语言模型中，句子是由一系列 token 组合而成的结构单元，体现了语言的上下文关联和整体语义。在 Enomic Language Model 中，细胞作为由基因表达量组合而成的功能单元，具有其内在的关联性和整体性。

We developed the GLM, which integrates data processing concepts from large language models \cite{zhang2010understanding,goldberg2017neural,amodio2019exploring} into its framework during the pretraining phase. Specifically, in single-cell transcriptomic data, each gene's expression value represents its unique role in cellular function and state, akin to the semantic representation of tokens in language models. Consequently, we redefined the basic units of single-cell transcriptomic data as conceptual structures within a language model framework. Each gene expression value is treated as the smallest data unit, analogous to a token in language models—a fundamental unit with independent semantics. Furthermore, we conceptualize a cell, defined by the expression values of all its genes, as equivalent to a sentence in language models. In large language models, sentences are structural units composed of a series of tokens that convey contextual relationships and overall semantics. Similarly, in the GLM, cells are functional units formed by the combination of gene expression values, reflecting their intrinsic associations and overall functions.

% 我们参考了 scGPT（Cui et al., 2024）中的方法，为基因字典中的每个基因分配了一个唯一的整数 ID。这使得我们能够将每个细胞的基因表达谱表示为一个向量：
% \[
% Cg^{[i]} = \left[ id(g_1^{[i]}), id(g_2^{[i]}), \ldots \right]
% \]

% 为了避免不同数据批次之间的尺度差异，我们采用了一种分箱技术。我们将同一批次的数据离散化为连续的区间 $[b_k, b_{k+1}] \quad \text{(其中 }k\in\{1,\ldots,n\}\text{)}$。因此，同一区间内的不同数据点最终被分配为相同的整数值，从而有效减轻了批次效应对模型训练的影响。
% 随后，我们通过传统的嵌入层将基因名称和基因表达值分别转换为 $emb_{\text{gene}}$ 和 $emb_{\text{value}}$。这些嵌入向量与其他条件信息一起构成了模型的输入。详细的数据处理工作流程如图 1a 所示。

We referenced the methods in scGPT \cite{scgpt} and assigned a unique integer ID to each gene in the gene dictionary. This allows us to represent the gene expression profiles of each cell as a vector
$$
Cg^{[i]} = \left[ id(g_1^{[i]}), id(g_2^{[i]}),\ldots \right](i\in\{1,\ldots,n\}).
$$
Here,  \( g \)  represents the gene, and \( i \) is the index of the gene,
To avoid scale differences between different batches of data, we adopted a binning technique. We discretized the data from the same batch into continuous intervals  $[b_k, b_{k+1}], \text{where }k\in\{1,\ldots,n\}\text{}$. Thus, different data points within the same interval were ultimately assigned the same integer value, effectively mitigating the impact of batch effects \cite{haghverdi2018batcheffects,tran2020benchmark} on model training.

\begin{figure*}[htbp]
    \centering
    \includegraphics[width=\textwidth]{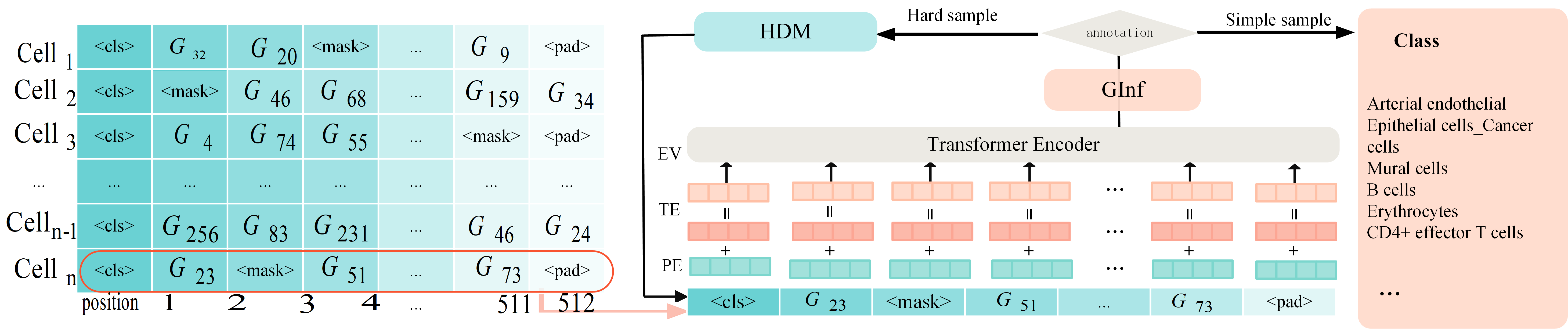}
    \caption{\label{结构} Schematic diagram of the model introduced structure, where PE is the Position Encoder, TE is the Token Embedding, and EV is the Embedding Vector.}
\end{figure*}

Subsequently, we converted the gene names and gene expression values into $emb_{gene}$ and $emb_{value}$ through the traditional embedding layer. Here, $emb_{gene}$ is analogous to the positional encoding information in large models. These embeddings, along with other conditional information, constitute the model's input. The detailed data processing workflow is shown in Figure \ref{结构}.

% 在预训练模型中，我们采用了 Transformer 模块和多头注意力机制，以增强模型捕捉不同基因之间相互关系的能力。这使得模型能够基于学习到的基因交互模式更好地执行细胞注释任务。
% 我们采用了常用的无监督掩码训练方法。具体而言，我们随机掩盖了 15\% 的基因表达值，并使用多层感知机网络预测被掩盖的值。我们通过最小化预测值与真实基因表达值之间的均方误差（MSE）损失来优化模型性能。

% 其中$emb_{gene}$类似大模型中的位置编码信息，将基因名称和基因表达值相加传入一个Transformer模块，我们采用了常用的无监督掩码训练方法。具体而言，我们随机掩盖了 15\% 的基因表达值，并使用MLP预测被掩盖的值。通过最小化预测值与真实基因表达值之间的均方误差（MSE）损失来优化模型性能，让模型学习不同细胞的特征。训练时针对long-tail分布的数据使用GInF损失，做到类别均衡，并且对困难样本(HDM)二次训练，从而提高模型在稀有类别上的性能表现。

% In the pre-training model, we adopted Transformer modules and the multi-head attention mechanism to enhance the model's ability to capture the interrelationships between different genes. We utilized the commonly employed unsupervised masked training approach. Specifically, we randomly masked 15\% of the gene expression values and employed a multilayer perceptron network to predict the masked values. We optimized the model's performance by minimizing the mean squared error (MSE) loss between the predicted and true gene expression values.

During the pre-training stage, we incorporated Transformer modules fortified with a multi-head attention mechanism to amplify the model's proficiency in recognizing the interdependencies among distinct genes. Capitalizing on the widely-used unsupervised masked training scheme, we intentionally masked 15\% of the gene expression values in a random fashion. Following this, we deployed a multilayer perceptron network to predict these obscured values. The optimization process served to further enhance the model's efficacy, particularly through the application of the GInf loss function and repetitive training on samples that were misclassified with high confidence.

\subsection{Long-Tailed Single-Cell Annotation}

%https://zhuanlan.zhihu.com/p/619079149
% 为了更好地探索单细胞时空组学，最基础的任务是对单细胞进行注释。基因语言模型在上游无监督阶段通过上下文掩码预训练的方式对海量无标签的单细胞数据进行知识的学习，使得基因语言模型能够准确的对单细胞mRNA数据进行准确的特征的抽取。在细胞注释的下游任务中我们将问题转化成为有监督的数据分类的问题。然而由于疾病的特殊性质，即便在病变的器官组织中，发生病变的细胞数量也是远远小于正常细胞的数量的。然而能够准确注释到疾病细胞确实对相关研十分重要的，这种长尾数据分布增加了细胞注释的难度。
To better explore single-cell spatiotemporal omics, the most fundamental task is to annotate single cells. In the upstream unsupervised stage, the gene language model learns from massive unlabeled single-cell data via context masking, enabling accurate feature extraction from mRNA data. In the downstream task of cell annotation, the problem is transformed into a supervised data classification problem. However, due to the unique nature of diseases, even within diseased tissues or organs, the number of diseased cells is significantly smaller than the number of normal cells. Accurately annotating diseased cells, however, is critically important for related research. This long-tailed data distribution further increases the difficulty of cell annotation.

\subsubsection{Traditional Definition: Cross-Entropy Loss}

% 我们将包含 $n$ 个样本的训练集表示为 $D = \{\mathbf{x}_i, y_i\}$，其中 $\mathbf{x}_i$ 表示第 $i$ 个细胞样本，$y_i$ 表示对应的注释。

% 对于传统工作的定义，这一过程可以有效地简化为一个分类深度学习任务，其中% 我们只需要利用经过良好训练的单细胞基因语言模型（Gene Language Model，GLM）提取细胞 $\mathbf{x}_i$ 的特征映射，然后将这些特征输入由 $\bm{\psi}$ 表示的分类器来计算将 $\mathbf{x}_i$ 分类到类别 $j$ 的预测 logit，公式如下所示（如公式~\ref{Z repertation} 所示）：

We denote the training set containing $n$ samples as $D = \{\mathbf{x}_i, y_i\}$, 
where $\mathbf{x}_i$ represents the $i$-th cell sample and $y_i$ represents its corresponding annotation.
For the definition of traditional tasks, this process can effectively be simplified into a classification deep learning task. the classifier of choice is the structurally simple VGG network \cite{vgg16}, where we only need to extract features using a well-trained single-cell GLM to extract the feature map of cell $\mathbf{x}_i$, and then input these features into a classifier represented by $\bm{\psi}$ to calculate the predicted logit of classifying $\mathbf{x}_i$ into category $j$, as shown in the following Equation ~\ref{Z repertation}:

\begin{equation}
\label{Z repertation}
z_{ij}= \psi_{j}(GLM(x_{i})).
\end{equation}

% 将细胞 $\mathbf{x}_i$ 分类到类别 $j$ 的预测概率通过将 logit $z_{ij}$ 输入 Softmax 函数定义，如公式~\ref{ce_prob} 所示：
The predicted probability of classifying $\mathbf{x}_i$ into category $j$, with $C$ denoting the total number of categories, is defined by passing the logit $z_{ij}$ through a Softmax function, as shown in Equation~\ref{ce_prob}:
\begin{equation}
\label{ce_prob}
% \tilde{\mathbf{p}}_j(\mathbf{X}_i; \bm{\psi}) = \frac{\exp(z_{iy_{i}})}{\sum_{j=1}^C \exp(z_{ij})}
\mathbf{p}_{j}(\mathbf{X}_i; \mathbf{\psi}) = \frac{\exp(z_{ij})}{\sum_{l=1}^C \exp(z_{ij})}.
\end{equation}
% 
% 然而这种基于CE 为loss 函数的大语言模型方法\cite会导致模型在训练过程中由于数据不均衡的问题把大量的注意力放在头部类别上，导致头部类别的过拟合以及尾部类别的欠拟合。从而导致对细胞注释的结果不尽人意。
Cross-entropy is commonly used as the loss function, as shown in Equation~\ref{ce}. However, this large language model approach based on the Cross-Entropy (CE) loss function \cite{scgpt,scbert,ICLR2024cellplm} tends to cause the model to focus a significant amount of attention on the head classes during training due to data imbalance issues. This results in overfitting of the head classes and underfitting of the tail classes, ultimately leading to suboptimal cell annotation results.

\begin{equation}
\label{ce}
    L_{\text{entropy}}(z_{ij}) = - \sum_{j=1}^{c} y_i \log(\mathbf{p}_{j}(\mathbf{X}_{i}; \mathbf{\psi})).
\end{equation}

\subsubsection{Gaussian Inflation Loss for Long-tail Distribution}

%进而,我们考虑到在人类疾病中，相较于正常细胞，患病细胞的数量相对较少，即使在受影响个体的器官内，患病细胞通常也仅占数据集的一小部分。由于感兴趣的细胞位于分布的尾部，网络难以专注于这些尾部类别，从而导致严重的长尾问题。

% 为缓解上述问题，一个可行的解决方案是减少头部类别对尾部类别施加的负样本梯度。因此，我们提出了高斯膨胀损失（Gaussian Inflation Loss，GInf Loss）作为解决方案。
% 通过参考借鉴一些reweight的方法 \cite{gaussian_cloud}\cite{seesaw}，我们基于CE loss为基础范式推导出GInf loss，见公式GInf。
 
Further, we consider that in human diseases, the number of diseased cells is relatively small compared to the number of normal cells, and even within the organs of affected individuals, diseased cells usually represent only a small fraction of the dataset.Since the cells of interest are located in the tails of the distribution, it is difficult for the network to focus on these tail categories, leading to a serious long-tail problem.
To alleviate the above-mentioned problem, a feasible solution is to reduce the negative sample gradients imposed by head classes on tail classes. Therefore, we propose Gaussian Inflation Loss (GInf Loss) as a solution.
By referring to and drawing inspiration from some reweighting methods \cite{gaussian_cloud,seesaw}, we derive the GInf loss based on the CE loss as the fundamental paradigm, as shown in Equation \ref{GInf}.
\begin{equation}
\label{GInf}
    % L_{\text{GInf Loss}}(z) = - \sum_{i=1}^{c} y_i \log(\hat{\sigma}_i), \tilde{\mathbf{p}}
\tilde{L}_{\text{GInf}}(z_{ij}) = - \sum_{j=1}^{c} y_i \log(\tilde{\mathbf{p}}_j(\mathbf{X}_{i}; \mathbf{\psi})).
\end{equation}

% GInf loss是通过对softmax函数进行优化，调节不同类别的超参数$N_{j}$，迫使网络对尾部类别更多的注意力，促进尾部类别的收敛。另外对${\psi}}$ 预测的logit变量进行修饰为$z_{ij}^{inf}$,见公式infp
The GInf loss is derived by optimizing the Softmax function and adjusting the hyperparameters $N_{j}$ of different categories, forcing the network to pay more attention to tail categories and promoting their convergence. Additionally, the logit variable ${\psi}$ predicted by the network is modified to $z_{ij}^{inf}$, as shown in Equation \ref{infp}.
\begin{equation}
\label{infp}
    % \hat{\sigma}_i = \frac{\exp({z_{i}^{inf}})} {\sum_{j \neq i}^{c} S_{ij} \exp({z_j}) + \exp({z_{i}^{inf}})}.\tilde{\mathbf{p}}
\tilde{\mathbf{p}}_j(\mathbf{X}_i; \mathbf{\psi}) = 
\frac{N_{j}\exp(z_{ij}^{inf})}
{\sum_{l \neq j}^{c} N_{l} \exp(z_{il}^{inf}) + N_{j} \exp(z_{ij}^{inf})}.
\end{equation}

% \begin{equation}
% \label{ce_prob}
% \tilde{\mathbf{p}}_j(\mathbf{X}_i; \bm{\psi}) = \frac{\exp(z_{ij})}{\sum_{l=1}^C \exp(z_{il})}
% \end{equation}

%为了解释其原理，我们绘制了图2，从图中可以观察到，我们在logit的特征空间，希望以高斯分布来扩增尾部类别单个个体所占的空间比重，来平衡头部类别在特征空间中所占的总体比重。

\begin{figure}[b] % 优先顶部和底部
\centering
\includegraphics[width=3.2in]{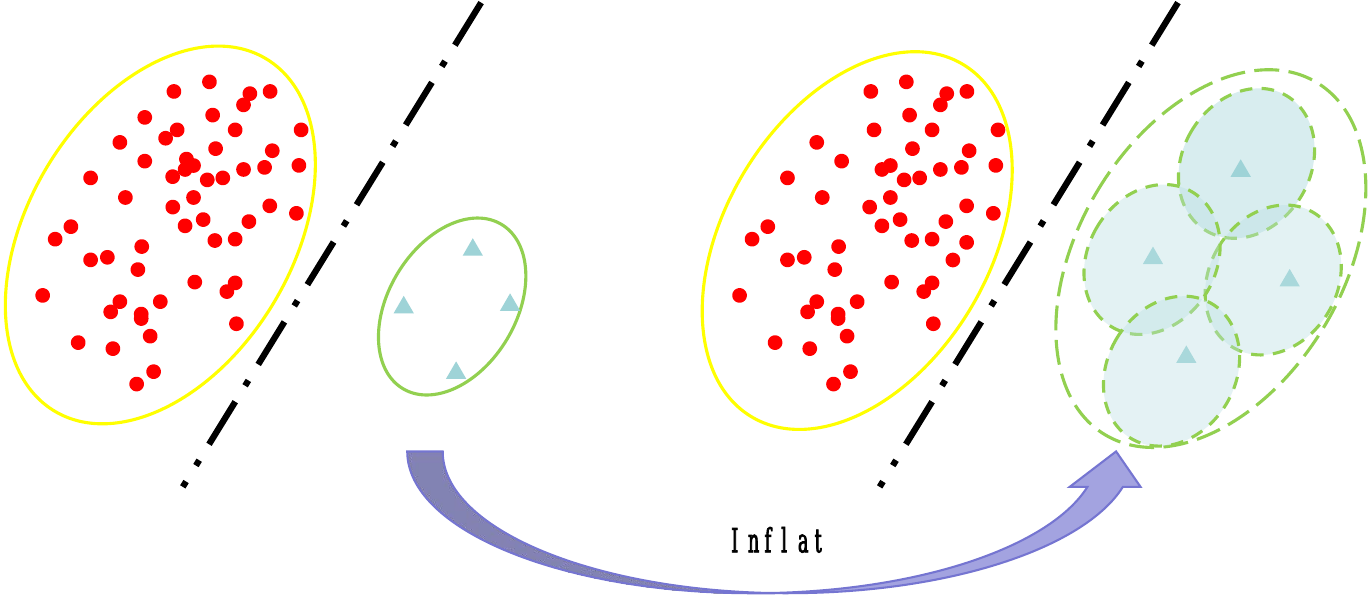}
\caption{Gaussian Inflation Expands Tail Classes.}
\label{inflat}
\end{figure}

The soul of our methed GInf Loss is $z_{ij}^{inf}$, which is Equation \ref{class_aware_loss}. To explain the underlying principle, we plot Figure 
 \ref{inflat}. From the figure \ref{inflat}, it can be observed that in the feature space of logits, we aim to use a Gaussian distribution to enlarge the spatial proportion occupied by individual instances of tail categories, thereby balancing the overall proportion occupied by head categories in the feature space.

% \begin{figure}[!htbp]
% \centering
% \includegraphics[width=3.2in]{inflat.pdf}
% % \caption{We added an inflation factor to the tail classes, increasing the feature space for these classes.}
% \caption{Gaussian Inflation Expands Tail Classes.}
% \label{inflat}
% \end{figure}

\begin{equation}
\label{class_aware_loss}
z_{ij}^{inf}=z_{ij}+\Delta \delta I.
\end{equation}

% 其中，$I \sim \mathcal{N}(\mu ,\Sigma )$ 是从高斯分布中采样的膨胀样本，$\mu$ 为均值向量，$\Sigma$ 为协方差矩阵，$\Delta \delta$ 是用于调整膨胀幅度的参数，我们将其控制为一个非常小的值。

% 类别膨胀因子与决策边界之间的关系建立了与类别相关的膨胀因子。为了保持类别膨胀因子的一致性，我们将膨胀因子设置为
where $ I \sim \mathcal{N}(\mu ,\Sigma )$ is infaction sample from Gaussian distribution and $\mu$ is the mean vector and $\Sigma$ is the convariance matrix, $\Delta \delta$ is a  parameter which is used to adjust the  amplitude of inflation and we control it to a very small number. The relationship between the inflation factor and the decision boundary establishes the inflation factor related to the class.
\begin{equation}
\label{class_aware_loss2}
 \Delta \delta=\log\ N_{max}-\log\ N_{j}.
\end{equation}
To keep the category inflation factor consistent, we set the inflation factor to be Equation \ref{class_aware_loss2}. Here, $N_{j}$  is the number of j classes freqent samples in the training set, the same as $N_{i}$.

\subsection{Hard Data Mining}

%%%%%%%%%%%%%% 下面应该引入难样本挖掘

%引入了高斯膨胀函数确实一定程度上缓解长尾数据分布带来的问题，但为了进一步提高模型的性能，我们提出了一种更为有效的策略，称为困难样本数据挖掘（Hard Data Mining, HDM），该方法通过专注于选择难以分类的类别进行训练，有效提升了模型对具有挑战性类别样本的区分能力。在困难样本数据挖掘中，\textit{HDM}被定义为那些尽管不是当前样本的真实类别，但在模型预测中得分较高的类别。换言之，HDM是指那些对模型而言容易产生混淆的类别，它们的预测得分接近甚至超过了真实类别的得分。为了选择这些困难样本，我们通过比较模型的输出值，确定得分较高但不匹配真实类别的类别作为重点关注对象。在分类任务中，假设总共有 $C$ 个类别，其中通过硬数据挖掘策略选取 $C_{\text{hard}}$ 个类别进行训练。具体而言，对于总样本 $\mathbf{x}_i$，我们构造一个难样本子集$\bm{\Omega}_i$，其中包含被选中的难分类类别。该集合由 $C_{\text{hard}}$ 个类别组成，这些类别通过从模型预测中排除真实类别后，选取得分最高的类别确定。通过这一策略，HDM能够使模型专注于难样本的学习，从而显著提升其对复杂数据的区分能力。对于样本 $\mathbf{x}_i$，包含选定类别输出的对应集合 $\bm{\Omega}_i$ 表示为：

% 在面对长尾数据分布问题时，高斯膨胀函数的引入为缓解这一问题提供了一定的帮助。然而，长尾分布问题的复杂性注定需要更有针对性的策略来进一步优化模型性能。因此，我们提出了一种名为困难样本数据挖掘（Hard Data Mining, HDM）的方法。与传统的训练方法不同，HDM 注重于挖掘那些对模型表现具有关键影响但训练难度较大的样本类别，从而促使模型更高效地学习复杂样本的特征分布。在 HDM 方法中，“困难样本”被定义为那些虽然不是当前样本的真实类别，但在模型预测中得分较高的类别。这些类别通常是模型最容易混淆的目标，因为它们的得分接近甚至有时可能超过真实类别的得分。这种混淆现象反映了模型当前的不足，因此专注于这些类别的学习能够有效提升模型的分类能力。具体而言，HDM 通过比较模型输出的预测得分，排除真实类别后，从得分最高的类别中动态选出难分类类别，作为训练的重点。假设总共有 $C$ 个类别，我们定义的困难类别集合 $\bm{\Omega}_ij$，是将原本为j类的$\mathbf{x}_ij$，被模型错误预测为其他类别，从c种错误预测的类别中依次取每个类别中logit得分前n高的困难样本组成，本实验中n取20。

% 这一过程不仅提高了模型对易混淆类别的区分能力，同时也让模型的训练资源得以高效利用。通过这样的设计，HDM 能够使模型将更多的关注点放在富有挑战性的样本上，而不是简单地优化容易分类的样本分布。这种策略的优越性在复杂、多样化的任务中尤为显著。困难类别集合 $\bm{\Omega}_i$ 的定义可以表示如下：对于任一样本 $\mathbf{x}_i$，由选定类别构成的困难类别集合 $\bm{\Omega}_i$ 可以表示为：

When dealing with the problem of long-tailed data distributions, the introduction of the Gaussian expansion function provides some relief. However, the complexity of the long-tailed distribution problem inevitably requires more targeted strategies to further optimize model performance. To this end, we propose a method called Hard Data Mining (HDM). Unlike traditional training methods, HDM focuses on identifying and learning from sample categories critical to model performance but challenging to train. This approach allows the model to efficiently learn the feature distributions of challenging samples. In the HDM method, ``hard samples" refer to categories that achieve high prediction scores in the model's outputs, despite not being the true category of the current sample. These categories are often the most confusing for the model, as their scores are close to or sometimes even exceed those of the true category. This confusion reflects the current shortcomings of the model; therefore, focusing on these categories can significantly improve the model's classification ability.

Specifically, HDM dynamically selects the hard-to-classify categories as training priorities by comparing the prediction scores output by the model, excluding the true category, and selecting the highest-scoring categories. Assuming there are $C$ categories in total, the defined hard category set $\bm{\Omega}_{ij}$ consists of hard samples for which the original class of $\mathbf{x}_{ij}$ was $j$ but were misclassified by the model into other categories. From the $C$ misclassified categories, the top $n$ hard samples with the highest logit scores in each category are selected to form the set. In this experiment, $n$ is set to 20.
This process not only enhances the model's ability to distinguish between confusing categories, but also ensures more efficient use of training resources. By this design, HDM allows the model to focus more on challenging samples rather than simply optimizing the distribution of easily classified samples. This strategy proves to be particularly advantageous in complex and diverse tasks.
The definition of the hard category set $\bm{\Omega}_i$ can be expressed as follows: for any sample $\mathbf{x}_i$, the hard category set $\bm{\Omega}_i$ composed of the selected categories can be expressed as \eqref{hard_select}.

\begin{equation}
\label{hard_select}
\bm{\Omega}_{i} = \bigcup_{l=1}^c \text{Top}_{n} \{x_{ij}, z_{il} \,|\, l \neq j \} \cup \{z_{ij} \},
\end{equation}
% 其中，$\text{TopHard}$ 表示选择具有最大值的 $C_{\text{hard}}$ 个样本。为了更好地适应长尾学习，我们以一种平衡的方式计算所选类别的概率，其表达形式为：
where $\text{Top}_{n}$ means selecting $C_{\text{hard}}$ examples with the largest values. 
In order to adapt better to long-tailed learning, we compute the probabilities of the selected categories in a balanced way, 
\begin{equation}
\label{hard_possibility}
    L^{\text{hard}}_{\text{GInf}}(z_{ij}) = -\sum_{j=1}^{c} \log \left( \frac{N_j \exp(z_{ij}^{\text{inf}})}{\sum_{z_{il} \in \bm{\Omega}_i} N_l \exp(z_{il})} \,\right),
\end{equation}
where $N_j$ represents the normalization term for category $j$. In the proposed GInf, this training strategy can be integrated into the previous process and executed synchronously. 

\begin{equation}
\label{total-loss}
    L_{\text{GInf}}(z_{ij}) = \tilde{L}_{\text{GInf}}(z_{ij}) + L^{\text{hard}}_{\text{GInf}}(z_{ij}).
\end{equation}
The total loss is ultimately expressed as the summation of GInf and the hard sample re-training terms.

% \begin{equation}
% \label{hard_possibility}
%     L^{\text{hard}}_{\text{GInf}}(z_{ij}) = -\sum_{j=1}^{c} \log \left( \frac{N_j \exp(z_{ij}^{\text{inf}})}{\sum_{z_{il} \in \bm{\Omega}_i} N_l \exp(z_{il})} \,\right)
% \end{equation}

% \begin{equation}
%     L^{\text{hard}}_{\text{GInf}}(z_{ij}) = -\sum_{j=1}^{c} \log \left( p^{hard}_{y_i} \left( \mathbf{x}_i; \boldsymbol{\psi} \right) \right)
% \end{equation}

% 在提出的GInf中，这种训练策略可以被融合到之前的过程中同步进行，其总损失表示为：

% \begin{equation}
% \label{total-loss}
%     L_{\text{GInf}}(z_{ij}) = \tilde{L}_{\text{GInf}}(z_{ij}) + L^{\text{hard}}_{\text{GInf}}(z_{ij})
% \end{equation}

\section{Experiments}

\subsection{Dataset}
% \begin{table}[h!]
%     \centering
%     \large % 调整字体大小为较大字体
%     % 缩小表格宽度
%     \resizebox{\linewidth}{!}{% 调整表格宽度为页面的 100%
%     \renewcommand{\arraystretch}{1.3} % 增加行距
%     \setlength{\tabcolsep}{9pt} % 减小列间距
%     \begin{tabular}{p{4.5cm}r|p{4.7cm}r}
%         \hline
%         \multicolumn{4}{l}{\textbf{Cell Types and Counts}} \\ \hline
%         \textbf{Cell Type} & \textbf{Count} & \textbf{Cell Type} & \textbf{Count} \\ \hline
%         Keratinocytes & 505131 & Langerhans cells & 15275 \\ 
%         T cells & 146555 & Melanocytes & 15269 \\ 
%         Fibroblasts & 117014 & Mast cells & 14769 \\ 
%         Endothelial cells & 97924 & Doublet & 12494 \\ 
%         Mural cells & 60523 & B cells & 10112 \\ 
%         Mononuclear phagocytes & 57495 & T cells, Cancer cells & 10008 \\ 
%         Sweat gland cells & 34130 & Schwann cells & 4636 \\ 
%         Melanocytes, Cancer cells & 27379 & Plasma cells & 3751 \\ 
%         Epithelial cells, Cancer cells & 16837 & Plasmacytoid dendritic cells & 2149 \\ 
%         Proliferating cells & 310 & Unassigned & 894 \\ 
%         Group2 innate lymphoid cells & 127 & Hair follicle cells & 413 \\ 
%         Hepatocytes & 76 & Merkel cells & 31 \\ 
%         Innate lymphoid cells & 32 & Erythrocytes & 18 \\ \hline
%     \end{tabular}%
%     }
%     \caption{Using skin tissue as an example, we provide an overview of the training data composition and highlight the long-tail distribution in the Celler-75 dataset.}
%     \label{tab:cell_counts}
% \end{table}
% \vspace{-0.1cm} % 添加间距

\vspace{-0.23cm} % 添加间距
\begin{table}[htbp]
\centering
\scriptsize % 进一步减小字体大小
\renewcommand{\arraystretch}{0.4} % 缩小行间距

\begin{tabular}{>{\centering\arraybackslash}m{1.2cm}|>{\centering\arraybackslash}m{1.4cm}|>{\centering\arraybackslash}m{1.8cm}|>{\centering\arraybackslash}m{1.74cm}}
\toprule

\textbf{Dataset}  & \textbf{Celler-75} & \textbf{MS} & \textbf{hPancreas} \\

\midrule Cell Num & 41,307,753         & 13,468      & 10,600            \\
\midrule Gene Num   & 21,292             & 3,000       & 3,000             \\
\midrule

& & & \\ [0.05ex]
%\multirow{\makecell{Celltype}} & \multicolumn{1}{c|}{\makecell{Full  20}} & \multirow{\makecell{18}} & \multirow{\makecell{13}} \\
%& \multicolumn{1}{c|}{\makecell{Sub  45}} & & \\
Celltype & \makecell{Full 20 \\ Sub 45} & 18 & 13\\                              
\midrule
Tissue      & 80                 & 3           & Pancreas-Only                 \\
\midrule Disease          & 75     & Multiple Sclerosis    & Pancreas-Related                 \\
\bottomrule
\end{tabular}
\caption{Comparison of datasets Celler-75, MS, and hPancreas.}
\label{tab:dataset_comparison}
\end{table}

\vspace{-0.55cm} % 添加间距

\begin{figure*}[ht]
    % \centering
    %\raggedright % 设置左对齐
   % \resizebox{0.5\textwidth}{!}{\includegraphics{pope.jpg}}
    \includegraphics[width=1.0\textwidth]{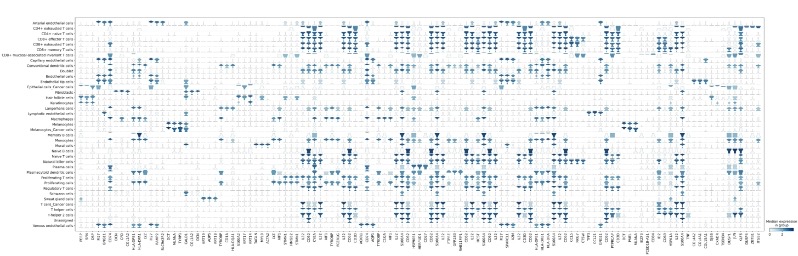}
    \caption{The bubble chart visualization illustrating gene expression levels and cell proportions.}
    \label{fig:dotplot}
\end{figure*}

\begin{table*}[t]
\centering
\resizebox{\textwidth}{!}{%
\begin{tabular}{c|c|ccc|ccc|ccc|ccc}
\toprule
% 第一行: 14 列
\multicolumn{1}{l|}{\textbf{Cell Type}} & \multicolumn{1}{l|}{\textbf{New Column}} & \multicolumn{3}{c|}{\textbf{SCBert}} & \multicolumn{3}{c|}{\textbf{SCgpt}} & \multicolumn{3}{c|}{\textbf{CellPLM}}  & \multicolumn{3}{c}{\textbf{Celler}} \\
% 第二行: 14 列
\multicolumn{1}{l|}{} & \multicolumn{1}{l|}{}  & \multicolumn{1}{c}{\textbf{f1-score}} & \multicolumn{1}{c}{\textbf{precision}} & \multicolumn{1}{c|}{\textbf{recall}} &\multicolumn{1}{c}{\textbf{f1-score}}  & \multicolumn{1}{c}{\textbf{precision}} & \multicolumn{1}{c|}{\textbf{recall}} & \multicolumn{1}{c}{\textbf{f1-score}} &\multicolumn{1}{c}{\textbf{precision}} & \multicolumn{1}{c|}{\textbf{recall}} & \multicolumn{1}{c}{\textbf{f1-score}} & \multicolumn{1}{c}{\textbf{precision}} & \multicolumn{1}{c}{\textbf{recall}}  \\

\midrule
% 数据行 1: Brain 分成两行
\multirow{2}{*}{Brain} & Parent-Classes & 0.727$^{\pm.012}$ & 0.705$^{\pm.203}$ & 0.782$^{\pm.007}$ & 0.919$^{\pm.032}$& 0.917$^{\pm.063}$ & 0.921$^{\pm.170}$ & 0.931$^{\pm.002}$ & 0.935 $^{\pm.006}$& 0.928$^{\pm.002}$ & \textcolor{red}{0.956$^{\pm.067}$}& \textcolor{red}{0.956$^{\pm.032}$} &\textcolor{red}{0.957$^{\pm.003}$} \\ 
                       & Subclasses                      &0.787$^{\pm.011}$ & 0.789$^{\pm.239}$& 0.806$^{\pm.025}$ & 0.802$^{\pm.036}$ & 0.827$^{\pm.041}$ & 0.777$^{\pm.104}$ & 0.779$^{\pm.126}$ & 0.794$^{\pm.081}$ & 0.765$^{\pm.037}$ & \textcolor{red}{0.890$^{\pm.026}$}&\textcolor{red}{ 0.886$^{\pm.051}$}&\textcolor{red}{ 0.895$^{\pm.113}$} \\

% 数据行 3: kidney
\midrule
\multirow{2}{*}{kidney} & Parent-Classes & 0.841$^{\pm.023}$ & 0.823$^{\pm.039}$ & 0.805$^{\pm.024}$  & 0.880$^{\pm.017}$ & 0.885$^{\pm.048}$ & \textcolor{red}{0.875$^{\pm.021}$} & \textcolor{red}{0.881$^{\pm.005}$} & \textcolor{red}{0.888$^{\pm.016}$} & 0.874$^{\pm.012}$ & 0.820$^{\pm.019}$ & 0.805$^{\pm.004}$ &0.835$^{\pm.016}$ \\ 
                       & Subclasses & 0.668$^{\pm.005}$ & 0.704$^{\pm.022}$ & 0.665$^{\pm.031}$ & 0.679$^{\pm.040}$ & 0.712$^{\pm.018}$ & 0.648$^{\pm.004}$ & 0.695$^{\pm.018}$ & 0.717$^{\pm.124}$ & 0.673$^{\pm.106}$ & \textcolor{red}{0.879$^{\pm.034}$} &\textcolor{red}{0.890$^{\pm.041}$}&\textcolor{red}{ 0.869$^{\pm.025}$}   \\

% 数据行 2: liver
\midrule
\multirow{2}{*}{liver} & Parent-Classes & 0.636$^{\pm.107}$ & 0.712$^{\pm.111}$ & 0.626$^{\pm.042}$ & 0.906$^{\pm.019}$ & 0.914$^{\pm.008}$ & 0.898$^{\pm.030}$ &  0.885$^{\pm.017}$  & 0.886$^{\pm.042}$  &  0.884$^{\pm.014}$ & \textcolor{red}{0.969$^{\pm.008}$} & \textcolor{red}{0.964$^{\pm.021}$}& \textcolor{red}{0.975$^{\pm.021}$} \\ 
                       & Subclasses                      & 0.618$^{\pm.031}$ & 0.628$^{\pm.004}$ & 0.641$^{\pm.021}$ & 0.789$^{\pm.067}$ & 0.806$^{\pm.034}$  & 0.773$^{\pm.029}$  & 0.759$^{\pm.038}$  & 0.785$^{\pm.033}$  &  0.735$^{\pm.029}$ &\textcolor{red}{0.840$^{\pm.024}$} &\textcolor{red}{0.868$^{\pm.016}$}& \textcolor{red}{0.813$^{\pm.015}$}   \\

\midrule
\multirow{2}{*}{Skin} & Parent-Classes  & 0.690$^{\pm.016}$ & 0.780$^{\pm.011}$ & 0.660$^{\pm.019}$ & \textcolor{red}{0.875$^{\pm.034}$}& \textcolor{red}{0.879$^{\pm.028}$} & 0.870$^{\pm.019}$ & \textcolor{red}{0.875$^{\pm.007}$} & 0.877$^{\pm.009}$ & \textcolor{red}{0.874$^{\pm.011}$} & 0.870$^{\pm.015}$ & 0.868$^{\pm.017}$ & 0.873$^{\pm.011}$  \\ 
                       & Subclasses                      &0.676$^{\pm.012}$& 0.674$^{\pm.009}$ & 0.724$^{\pm.023}$
                      & 0.630$^{\pm.043}$ & 0.666$^{\pm.028}$ & 0.596$^{\pm.030}$ & 0.685$^{\pm.024}$ & 0.706$^{\pm.018}$ & 0.666$^{\pm.014}$ & \textcolor{red}{0.860$^{\pm.024}$} & \textcolor{red}{0.862$^{\pm.031}$} & \textcolor{red}{0.859$^{\pm.035}$}   \\
% 数据行 4: lung
% \midrule
% \multirow{2}{*}{lung} & \multirow{1}{*}{Parent-Classes}  & 0 & 0 & 0 & 0.988& 0.953 & 0.942 & 0. & 0. & 0. & 0.984 & 0.933 & 0.947\\ 
%                        & \multirow{1}{*}{Subclasses}                        & & 0.895 & 0.894 & 0.913 & 0.895 & 0.894 & 0.900 & 0.894&0.895 & 0.918 & 0.901 & 0.905    \\
% \midrule
% \multirow{2}{*}{Cortex} & \multirow{1}{*}{Parent-Classes}  & 0 & 0 & 0 &0.977 & 0.936 & 0.957 & 0. & 0. & 0. & 0.984 &0.964 & 0.975 \\ 
%                        & \multirow{1}{*}{Subclasses}                         & & 0.804 & 0.820 & 0.972 & 0.804 & 0.820 & 0.813 & 0.805 & 0.807 &  0.977& 0.906 & 0.866  \\

% \midrule
% \multirow{2}{*}{Breast} & \multirow{1}{*}{Parent-Classes}  & 0 & 0 & 0 & 0.976& 0.890 & 0.912 & 0. & 0. & 0. & 0.969 & 0.897 & 0.912  \\ 
%                        & \multirow{1}{*}{Subclasses}                          & & 0.870 & 0.827 
%                        & 0.869 & 0.820 & 0.837 & 0.818 & 0.821 & 0.815 & 0. & 0. &0.   \\

\bottomrule
\end{tabular}
}

\caption{Performance comparison between Celler and other advanced approaches (i.e., SCBert, SCGpt, CellPLM) across different tissues (i.e., brain, kidney, liver, skin).}

\label{tab:comparison in celler}
\end{table*}
% \vspace{-0.3cm} % 添加间距

\subsubsection{Multiple Sclerosis (MS) and hPancreas dataset}

% MS 和 hPancreas 数据集都是具有公开获取渠道的，针对特定人类疾病的数据资源。
% hPancreas 数据集专注于人类胰腺的研究。它包含了丰富的细胞转录组基因表达数据，训练集中收录了 10600 条数据，测试集中则包含了 4218 条数据。这些数据广泛应用于胰腺相关疾病的机制研究、诊断标志物的发现以及治疗策略的开发。

% 另一方面，MS 数据集聚焦于多发性硬化症的研究.该数据集包含来自患者和健康对照组的多模态数据，用于探究疾病的病理机制、筛选潜在生物标志物。具体来说，MS 数据集包括 13468 条训练数据和 7844 条测试数据，为多发性硬化症在基础研究和临床转化中提供了强有力的支持。

The Multiple Sclerosis (MS) \cite{MSdataset} and hPancreas \cite{hPancreas-dataset} datasets are publicly accessible resources specifically designed for studying certain human diseases. The detail of these two public dataset was show in Table \ref{tab:dataset_comparison}.
The hPancreas dataset focuses on the study of the human pancreas. It contains a wealth of single-cell transcriptomic gene expression data, with 10,600 samples in the training set and 4,218 samples in the test set. These data are widely used in research on pancreatic-related disease mechanisms, the discovery of diagnostic biomarkers, and the development of therapeutic strategies.

On the other hand, the MS dataset is dedicated to the study of multiple sclerosis. This dataset includes multimodal data from both patients and healthy control groups, aimed at exploring the pathological mechanisms of the disease and identifying potential biomarkers. Specifically, the MS dataset consists of 13,468 training samples and 7,844 test samples, providing strong support for both basic research and clinical translation efforts in the field of multiple sclerosis.

% % MS 和 hPancreas 数据集都是具有公开获取渠道的，针对特定人类疾病的数据资源。
% % hPancreas 数据集专注于人类胰腺的研究，提供了用于基础科学研究、生物医学应用以及临床分析的重要数据支持。它包含了丰富的细胞转录组基因表达数据，训练集中收录了 10600 条数据，测试集中则包含了 4218 条数据。这些数据广泛应用于胰腺相关疾病的机制研究、诊断标志物的发现以及治疗策略的开发。

% % 另一方面，MS 数据集聚焦于多发性硬化症的研究，这是一个复杂的神经系统疾病数据资源库。该数据集包含来自患者和健康对照组的多模态数据，用于探究疾病的病理机制、筛选潜在生物标志物、开发诊断工具以及寻找治疗方法。具体来说，MS 数据集包括 13468 条训练数据和 7844 条测试数据，为多发性硬化症在基础研究和临床转化中提供了强有力的支持。

% The MS \cite{MSdataset} and hPancreas \cite{hPancreas-dataset} datasets are publicly accessible resources specifically designed for studying certain human diseases.
% The hPancreas dataset focuses on research related to the human pancreas, providing critical data support for basic scientific research, biomedical applications, and clinical analysis. It contains a wealth of cell transcriptome gene expression data, with 10,600 samples in the training set and 4,218 samples in the test set. These data are widely used in studies on the mechanisms of pancreas-related diseases, the discovery of diagnostic biomarkers, and the development of therapeutic strategies.

% On the other hand, the MS dataset centers on research into multiple sclerosis, a complex neurological disease. This dataset includes multimodal data from both patients and healthy controls, which are utilized for exploring disease mechanisms, identifying potential biomarkers, developing diagnostic tools, and discovering treatment strategies. Specifically, the MS dataset comprises 13,468 training samples and 7,844 test samples, providing robust support for both basic research and clinical translation related to multiple sclerosis.

\subsubsection{Celler-75 dataset}

% Celler-75 是我们自主构建的高维单细胞数据集，专注于 70 个主要人体器官中的 75 种特定人类疾病，包括癌症、阿尔茨海默病等多种重大疾病。相较于公开的单器官数据集 MS 和 hPancreas，Celler-75 具有显著优势，其数据规模巨大，整合了超过 4000 万个单细胞数据和 900 万多个基因表达谱，包含的细胞数量是公共数据集的 3000 倍左右。由于数据集的规模提升，在细胞注释的类别标签制作中，我们更加细化了分类的维度，有目的的区分子类和父类两个层级的标签，具体会在下一节metric展开介绍。如此大规模、高质量的单细胞数据，为大规模基因模型的训练和微调提供了坚实基础，并显著提升了 Celler 在细胞类型广度和注释精度上的表现。该数据集涵盖了多种癌症类型及其他复杂的细胞病理状态，为全面解析疾病相关基因表达在细胞层面的异质性和特异性提供了强有力的支持。

Celler-75 is a high-dimensional single-cell dataset independently constructed by us, focusing on 75 specific human diseases across 70 major human organs, including major diseases such as cancer and Alzheimer’s disease. As shown in Table\ref{tab:dataset_comparison}, when compared with publicly available single-organ datasets like MS and hPancreas, Celler-75 demonstrates significant advantages due to its massive scale, which integrates over 40 million single-cell data points and more than 9 million gene expression profiles. The number of cells included in Celler-75 is approximately 3,000 times that of public datasets. Due to the increased scale of the dataset, we have refined the classification dimensions in the creation of cell annotation category labels, intentionally distinguishing between subclass and parent class levels. The specifics will be elaborated in the next section on metrics. Such a large-scale, high-quality single-cell dataset provides a solid foundation for the training and fine-tuning of large-scale gene models, significantly enhancing Celler's performance in terms of cell type diversity and annotation accuracy. This dataset encompasses various types of cancers and other complex cellular pathological states, providing robust support for comprehensively analyzing the heterogeneity and specificity of disease-related gene expression at the cellular level.
\vspace{-0.1cm} % 添加间距
% 另外，70种器官类型和以及每一种器官的数据分布及在整体数据集所占的比重被可视化的展示在 Supplementary Figure 1中。不仅如此，75种疾病的类别信息被逐个列出在Supplementary Table 1.
Additionally, the data distribution of 70 organ types and their proportions in the overall dataset are visualized in Supplementary Figure 1. Furthermore, the category information of 75 diseases is listed individually in Supplementary Table 1.
% % 表格1以皮肤组织中不同细胞类型及其对应的数量分布展示了celler40数据长尾的情况。皮肤组织细胞的数据分布呈现明显的层次结构。高频细胞类型如角质形成细胞（505,131）和T细胞（146,555）.中频细胞类型如内皮细胞和平滑肌细胞（57,495-97,924），主要分布在真皮层或参与循环调节。低频细胞类型如黑色素细胞和浆细胞（几千到几万），在特定功能和疾病中具有重要性。稀有细胞类型如梅克尔细胞（31）和红细胞（18），虽数量极少，但具有独特的生物学和医学意义。
% % 我们将更详细的celler-75的数据信息展示在补充工作中.

% Table \ref{} illustrates the long-tail distribution of the Celler40 dataset through the different cell types in skin tissue and their corresponding quantities.The distribution of skin tissue cells demonstrates a clear hierarchical structure. High-frequency cell types, such as keratinocytes (505,131) and T cells (146,555). Medium-frequency cell types, such as endothelial cells and mural cells (57,495–97,924), are primarily located in the dermis or involved in circulation regulation. Low-frequency cell types, such as melanocytes and plasma cells (ranging from thousands to tens of thousands), play significant roles in specific functions and diseases. Rare cell types, such as Merkel cells (31) and erythrocytes (18), are extremely scarce but possess unique biological and medical significance.We provide more detailed information about the Celler-75 dataset in the supplementary work.

\subsection{Metric}
% 在单细胞转录组学研究中，细胞类型的分类通常包括父类（Parent/Full Class）和亚类（Subclass），共同构成了细胞分类的层次结构。在我们的私有数据集 Celler-40 中，对于父类和子类两个层级，都提供了具有生物学意义的标签。

% “父类”指的是具有相似功能或共同发育起源的细胞群体，这些类别是基于全局基因表达特征定义的。而“子类”是父类的更具体表现，代表了更精细的细胞类型或状态，具有独特的功能特性以及标志性基因的表达特征。

% 例如，在“免疫细胞”这个父类下，其子类可能包括 T 细胞、B 细胞以及自然杀伤细胞（NK 细胞）。因此，如果某个细胞被鉴定为 T 细胞，其父类标签为“免疫细胞”，子类标签则为 “T 细胞”。

% 在这一层次结构中，准确注释细胞类型对于理解细胞多样性和功能异质性至关重要。为实现这一目标，需要利用稳健的评估指标来衡量预测注释与真实细胞类型的匹配程度。我们在此使用三个关键指标来评估细胞注释方法的性能：

% F1 Score：兼顾 Precision 和 Recall，在标注的准确性和覆盖性之间找到平衡，是细胞注释性能的综合评价指标.
% 精确率（Precision）：预测特定细胞类型（例如免疫细胞）的准确性，通过减少假阳性来确保细胞不被错误标记。
% 召回率（Recall）：衡量正确识别属于特定类型细胞的能力，通过降低假阴性来提高稀有细胞类型的注释准确性，这一点尤为重要。
In single-cell transcriptomics studies, the classification of cell types typically includes Parent Classes and Subclasses, which together form a hierarchical structure of cell categorization. In our private dataset, Celler-75, biologically meaningful labels for both the Parent Class and Subclass levels are provided.
Parent Classes refer to groups of cells with similar functions or shared developmental origins, and these categories are defined based on global gene expression characteristics. Subclasses, on the other hand, are more specific manifestations of Parent Classes, representing finer-grained cell types or states with distinct functional characteristics and expression of marker genes.
For example, under the Parent Class of immune cells, Subclasses may include T cells, B cells, and natural killer (NK) cells. Thus, if a cell is identified as a T cell, its Parent Class label would be immune cells, and its Subclass label would be T cells.

Accurately annotating cell types across this hierarchical structure is critical for understanding cellular diversity and functional heterogeneity. To achieve this, robust evaluation metrics are required to assess how well the predicted annotations align with the true cell types. Here, we employ three key metrics to evaluate the performance of cell annotation methods:
\textit{F1 Score}: Balances Precision and Recall, serving as a comprehensive evaluation metric for cell annotation performance by finding a trade-off between annotation accuracy and coverage. \textit{Precision}: the accuracy of predicting specific cell types (e.g., immune cells) by minimizing false positives, ensuring cells are not incorrectly labeled. \textit{Recall}: measures the ability to correctly identify cells belonging to a particular type by reducing false negatives, which is especially important for annotating rare cell types.
% \begin{itemize}
%      \item \textbf{accuracy:} a visual measure of the model's performance on the entire dataset, reflecting the model's average predictive power across all categories.
%     \item \textbf{Precision:} the accuracy of predicting specific cell types (e.g., immune cells) by minimizing false positives, ensuring cells are not incorrectly labeled. 
%     \item \textbf{Recall:} measures the ability to correctly identify cells belonging to a particular type by reducing false negatives, which is especially important for annotating rare cell types.
   
% \end{itemize}

\subsection{Evaluation}
\subsubsection{Cell Annotation Performance on Celler-75}

% 在横向对比实验中，我们选取了三项发表在 \textit{Nature Methods} 期刊上的 SOTA 方法作为对比实验，包括 CellPLM\cite{cellPLM}、scBERT\cite{scbert} 和 scGPT\cite{scgpt}，实验结果展示于表2中。对比实验的训练数据集选取自四个器官——肝脏、肾脏、皮肤和大脑，每种细胞类型根据两种粒度进行评估：Parent-Classes和 Subclasses。数据集总规模为 80 万条，其中 60 万条用于微调训练，20 万条用于测试。其中，每个器官的测试数据包含的细胞类别分布为 Brain：52 个亚类，32 个父类；Skin：47 个亚类，21 个父类；Liver：48 个亚类，24 个父类；Kidney：59 个亚类，32 个父类。实验结果表明，Parent-Classes 的分类性能普遍高于 Subclasses，这表明细粒度的子类分类任务对算法提出了更高的挑战。如表2所示，Celler 方法在 \textit{f1-score}、\textit{precision} 和 \textit{recall} 三个指标上表现出显著优势，其性能指标（标为红色的数值）普遍高于其他方法，次优结果则用蓝色标记。然而，在某些场景下（例如 Kidney 的 Parent-Class 和 Skin 的 Parent-Class 分类任务），Celler 的性能表现存在一定波动。

In the horizontal comparison experiments, we selected three SOTA methods published in the journal \textit{Nature Methods} as comparative experiments, including CellPLM \cite{ICLR2024cellplm} , scBERT \cite{scbert}, and scGPT \cite{scgpt}. The experimental results are presented in Table \ref{tab:comparison in celler}. The training dataset for the comparison experiments was sourced from four organs—liver, kidney, skin, and brain. Each cell type was assessed at two levels of granularity: Parent-Classes and Subclasses. The dataset contains a total of 800,000 samples, with 600,000 used for fine-tuning and 200,000 for testing. The distribution of cell categories in the test data for each organ is as follows: Brain: 52 subclasses, 32 parent classes; Skin: 47 subclasses, 21 parent classes; Liver: 48 subclasses, 24 parent classes; Kidney: 59 subclasses, 32 parent classes. The experimental results indicate that the classification performance of Parent-Classes is generally better than that of Subclasses, suggesting that finer-grained subclass classification tasks pose greater challenges to the algorithms. As shown in Table \ref{tab:comparison in celler}, the Celler method demonstrates significant advantages in the \textit{F1-score}, \textit{precision}, and \textit{recall} metrics, with its performance indicators (values marked in red) generally surpassing other methods. The second-best results are marked in blue. However, in certain scenarios (e.g., the Parent-Class classification tasks for Kidney and Skin), the performance of the Celler method shows some fluctuations.

\subsubsection{Cell Annotation Performance on Open Access Dataset}

\begin{table}[ht]
\centering
\resizebox{\columnwidth}{!}{%
\begin{tabular}{lcccc}
\toprule
\textbf{Method} & \multicolumn{2}{c}{\textbf{MS}} & \multicolumn{2}{c}{\textbf{hPancreas}} \\
\cmidrule(lr){2-3} \cmidrule(lr){4-5}
 & \textbf{F1 ($\uparrow$)} & \textbf{Precision ($\uparrow$)} & \textbf{F1 ($\uparrow$)} & \textbf{Precision ($\uparrow$)} \\
\midrule
CellTypist & 0.667$^{\pm.002}$ & 0.693$^{\pm.001}$ & 0.708$^{\pm.023}$ & 0.736$^{\pm.025}$ \\
ACTINN & 0.628$^{\pm.012}$ & 0.634$^{\pm.009}$ & 0.705$^{\pm.005}$ & 0.709$^{\pm.006}$ \\
SingleCellNet & 0.637$^{\pm.001}$ & 0.700$^{\pm.001}$ & 0.739$^{\pm.006}$ & 0.761$^{\pm.004}$ \\
TOSICA* & 0.578 & 0.664 & 0.656 & 0.661 \\
\midrule
scBERT  & 0.599$^{\pm.001}$ & 0.604$^{\pm.004}$ & 0.685$^{\pm.003}$ & 0.699$^{\pm.007}$ \\
scGPT   & 0.703$^{\pm.002}$ & 0.729$^{\pm.002}$ & 0.718$^{\pm.003}$ & 0.735$^{\pm.001}$ \\
\midrule
\textit{Celler}  & 0.799$^{\pm.004}$ & 0.841$^{\pm.002}$ & 0.767$^{\pm.010}$ & 0.755$^{\pm.010}$ \\
% \midrule
% \textit{Celler}  & \textbf{0.766$^{\pm.007}$} & \textbf{0.803$^{\pm.008}$} & \textbf{0.749$^{\pm.010}$} & \textbf{0.753$^{\pm.010}$} \\
\bottomrule
\end{tabular}%
}
%\caption{\textbf The results of cell type annotation on the MS and hPancreas datasets.  \textbf{*} indicates results directly taken from \cite{ICLR2024cellplm}.}
\caption{\textbf{The results of cell type annotation on the MS and hPancreas datasets.} \textbf{*} indicates results directly taken from CellPLM.}
\label{tab:task3-results}
\end{table}

% 从表格结果可以看出，\textbf{HiCeller} 在 MS 和 hPancreas 两个数据集上均表现最佳，展现了其在分类任务上的优越性。相比之下，\textbf{SingleCellNet} 在 hPancreas 数据集上的 Precision 达到了 $0.761 \pm 0.004$，紧随其后，但整体 F1 值仍低于 HiCeller。 \textbf{scGPT} 在MS数据集上表现较优。其他方法如 \textbf{CellTypist}、\textbf{ACTINN} 表现中规中矩，而 \textbf{TOSICA} 在两个数据集上的结果令人不太满意。综上所述 HiCeller 在处理这些任务时具有更强的鲁棒性和准确性。
We follow the suggestion of CellPLM \cite{ICLR2024cellplm} and GenePT \cite{chen2023genept} to include hPancreas \cite{hPancreas-dataset} and
Multiple Sclerosis (MS) \cite{MSdataset} datasets. As can be see table \ref{tab:task3-results}, Celler outperforms CellTypist \cite{dominguezconde2022crosstissue}, ACTINN \cite{ma2020actinn}, SingleCellNet \cite{tan2019singlecellnet}, TOSICA \cite{hPancreas-dataset}, scBERT \cite{scbert}, and scGPT \cite{scgpt}, achieving the highest performance on both the MS and hPancreas datasets, demonstrating its superiority in classification tasks. In comparison, SingleCellNet achieves a Precision of $0.761 \pm 0.004$ on the hPancreas dataset, ranking closely behind, but its overall F1 score is still lower than Celler. ScGPT performs well on the MS dataset. Other methods, such as CellTypist and ACTINN, show moderate performance, while TOSICA produces less satisfactory results on both datasets. In summary, Celler demonstrates stronger robustness and accuracy in handling these tasks.
% 这些结果表明，HiCeller 能够识别基因表达模式的底层特征以及远程基因-基因依赖关系，并实现对细胞类型特异性全局信息的全面高级表示.
These results indicate that, Celler is able to identify the underlying features of gene expression patterns as well as remote gene-gene dependencies, and achieve a comprehensive high-level representation of cell type-specific global information.

\subsection{Variation in Gene Expression Characteristics across Cell Types.}
% 图4展示了不同 基因（横轴） 在不同 细胞类型（纵轴） 中的表达模式，以及表达的显著性和广泛性。通过气泡的大小和颜色深浅反映基因的表达水平。在免疫细胞中，如 T 细胞和巨噬细胞，某些基因表现出显著表达。例如，CD3D、CD3E、CD4 和 HLA-DRB1 在 CD4+ T 细胞（如 CD4+ 枯竭 T 细胞和 CD4+ 初始 T 细胞）中表达显著，而 CD8A 和 CD8B 则特异性表达于 CD8+ T 细胞。同样，巨噬细胞中基因 CD68 和 HLA-DRB1 高度表达。在内皮细胞（如动脉内皮细胞和毛细血管内皮细胞）中，CDH5 和 VWF 是标志性基因，表明其与血管生成和维持密切相关。上皮细胞和癌细胞中，KRT19 和 KRT7 表现突出，分别反映上皮细胞特性及肿瘤相关表达变化。干细胞（如 Stem/B 细胞）中，PROM1 和 SOX2 与其增殖和分化功能密切相关。未分类或未分化细胞中显示部分基因表达水平较低，而特定细胞类型如角质形成细胞和朗格汉斯细胞中，KRT14/KRT5 和 CD207 作为典型标志，反映了这些细胞的独特生物学功能。
Figure \ref{fig:dotplot} illustrates the expression patterns of different genes (horizontal axis) across various cell types (vertical axis), as well as the significance and breadth of their expression. Gene expression levels are represented by the size and color intensity of the bubbles. In immune cells, such as T cells and macrophages, certain genes show prominent expression. For example, CD3D, CD3E, CD4, and HLA-DRB1 are significantly expressed in CD4+ T cells (e.g., CD4+ exhausted T cells and CD4+ naive T cells), while CD8A and CD8B are specifically expressed in CD8+ T cells. Likewise, genes such as CD68 and HLA-DRB1 show high expression levels in macrophages. In endothelial cells (e.g., arterial and capillary endothelial cells), CDH5 and VWF serve as signature genes, indicating their close association with angiogenesis and vascular maintenance. In epithelial and cancer cells, KRT19 and KRT7 are highly expressed, reflecting epithelial cell characteristics and tumor-related expression changes. In stem cells (e.g., Stem/B cells), PROM1 and SOX2 are closely linked to proliferation and differentiation functions. Unassigned or undifferentiated cells exhibit lower expression levels for certain genes, while specific cell types, such as keratinocytes and Langerhans cells, show distinct markers like KRT14/KRT5 and CD207, highlighting their unique biological functions.

\subsection{Ablation Experiments}
\begin{table}[h]
    \centering
     \small
    \begin{tabular}{lccc}
        \hline
        \textbf{Methods} & \textbf{Precision} & \textbf{Recall} & \textbf{F1-score} \\
        \hline
        CE Loss          & 0.910$^{\pm.004}$  & 0.826$^{\pm.006}$  & 0.797$^{\pm.007}$ \\
        Focal Loss       & 0.919$^{\pm.007}$  & 0.862$^{\pm.009}$  & 0.821$^{\pm.011}$ \\
        
        Ride Loss       & 0.921$^{\pm.005}$  & 0.851$^{\pm.009}$  & 0.877$^{\pm.010}$ \\
        Ours             & 0.951$^{\pm.003}$  & 0.886$^{\pm.012}$  & 0.895$^{\pm.009}$ \\
        \hline

    \end{tabular}
     \caption{Ablation study on different loss functions.}
         \label{tab:loss_comparison}
\end{table}
   \vspace{-0.1cm} % 添加间距
% 为了验证我们提出的扩张损失在长尾分布数据集上的有效性，我们在人类脑部疾病数据集上进行了消融实验。实验分别与知名的长尾损失函数（如 Focal Loss 和 BCL Loss）以及未考虑长尾分布的传统交叉熵损失（CE Loss）进行了对比。结果表明，我们的方法在多个指标上均取得了显著提升，尤其是在平衡精确率与召回率的 F1 分数方面，相较于 CE Loss 提高了近 10%。这充分证明了我们的方法在处理长尾分布分类任务时的卓越性能。具体实验结果详见下表：
To validate the effectiveness of our proposed expansion loss on long-tailed datasets, we conducted an ablation study on a human brain disease dataset. The experiment compared our method with well-known long-tailed loss functions (e.g., Focal Loss \cite{focalloss} and RIDE Loss \cite{kumar2021a2ride} ) as well as the traditional cross-entropy loss (CE Loss) \cite{cross-entropy} from that does not consider long-tailed distributions. The results demonstrate that our method achieved significant improvements across multiple metrics, particularly in balancing precision and recall with the F1-score, showing an increase of nearly 10\% compared to CE Loss. This strongly validates the superior performance of our method in handling classification tasks on long-tailed distributions. Detailed results are presented in the Table\ref{tab:loss_comparison}.

\section{Conclusion}
Our research results show that Celler performs extremely well when facing the public datasets MS, hPancreas, and our large-scale private dataset Celler-75. It improves the accuracy of identifying rare cell types by over 10\% compared with the scBert model, and its overall F1 Score is also significantly better than other benchmark models, highlighting its excellent generalization capability and practical application value. Efficient single-cell data annotation is of great significance for disease diagnosis, personalized medicine, and the discovery of biomarkers. The introduction of Celler not only breaks the limitations of existing annotation methods in handling long-tail distribution data but also provides strong tool support for the integrated analysis of complex multi-omics data in the future.

\section*{Acknowledgments}
This research received generous support from the National Natural Science Foundation of China (Grant No. 62076092) pertaining to the project titled "Intelligent Speech Dialogue System Response Generation for Affective Computing" spanning the duration 2021-2024.

% \nocite{cao2020scsa}
% \nocite{huang2020evaluation}
% \nocite{mofitt2018molecular}
% \nocite{zhang2019scina}
% \nocite{pliner2019supervised}
% \nocite{grabski2022probabilistic}
% \nocite{haghverdi2018batcheffects}
% \nocite{eisenstein2020single}
% \nocite{tran2020benchmark}
% \nocite{ganin2015unsupervised}
% \nocite{ceglia2023identification}
% \nocite{kim2020single}
% \nocite{paszke2019pytorch}
% \nocite{wolf2018scanpy}
% \nocite{danese2021episcanpy}
% \nocite{fang2023gseapy}
% \nocite{wolf2018scanpy}
% \nocite{danese2021episcanpy}
% \nocite{fang2023gseapy}

% \nocite{macparland2018single}
% \nocite{litvinukova2020cells}
% \nocite{tucker2020transcriptional}
% \nocite{lukassen2020sarscov2}
% \nocite{he2020single}
% \nocite{zhang2019cellmarker}
% \nocite{kimmel2021semi}
% \nocite{hao2021integrated}
% \nocite{hwang2019humannet}
% \nocite{liu2021transformers}
% \nocite{yun2019graph}
% \nocite{mcdavid2013data}
% \nocite{goldberg2017neural}
% \nocite{zhang2010understanding}
% \nocite{kharchenko2014bayesian}
% \nocite{hwang2019humannet}
% \nocite{liu2021transformers}
% \nocite{yun2019graph}
% \nocite{mcdavid2013data}

% \nocite{goldberg2017neural}
% \nocite{zhang2010understanding}
% \nocite{kharchenko2014bayesian}
% \nocite{amodio2019exploring}
% \nocite{brbic2020mars}
% \nocite{tarashansky2021mapping}
% \nocite{rives2021biological}
% % \nocite{elnaggar2022prottrans}
% \nocite{lin2023evolutionary}

%\bibliographystyle{named}
\bibliography{ijcai25}

\end{document}